\documentclass[twocolumn,tighten,times]{aastex61}

\usepackage{graphicx}
\usepackage{xspace}
\usepackage{amssymb}
\usepackage{amsmath}

\shorttitle{Death spiral of a binary stellar merger}
\shortauthors{Pejcha et al.}

\newcommand{\mdot}{\ensuremath{\dot{M}}}
\newcommand{\pdot}{\ensuremath{\dot{P}}}
\newcommand{\teff}{\ensuremath{T_{\rm eff}}}
\newcommand{\msun}{\ensuremath{M_{\odot}}}
\newcommand{\lsun}{\ensuremath{L_{\odot}}}
\newcommand{\myr}{\ensuremath{\msun\,{\rm yr}^{-1}}}
\newcommand{\ltwo}{L\ensuremath{_2}\xspace}

\begin{document}

\title{Pre-explosion spiral mass loss of a binary star merger}

\correspondingauthor{Ond\v{r}ej Pejcha}
\email{pejcha@utf.mff.cuni.cz}

\author{Ond\v{r}ej Pejcha}
\altaffiliation{Lyman Spitzer Jr.\ Fellow}
\affiliation{Institute of Theoretical Physics, Faculty of Mathematics and Physics, Charles University in Prague, Czech Republic}
\affiliation{Department of Astrophysical Sciences, Princeton University}
\affiliation{Kavli Institute of Theoretical Physics, University of California, Santa Barbara}

\author{Brian D. Metzger}
\affiliation{Columbia Astrophysics Laboratory, Columbia University, New York, NY 10027, USA}

\author{Jacob G. Tyles}
\affiliation{Department of Astrophysical Sciences, Princeton University}

\author{Kengo Tomida}
\affiliation{Department of Earth and Space Science, Osaka University}

\begin{abstract}
Binary stars commonly pass through phases of direct interaction which result in the rapid loss of mass, energy, and angular momentum.  Though crucial to understanding the fates of these systems, including their potential as gravitational wave sources, this short-lived phase is poorly understood and has thus far been unambiguously observed in only a single event, V1309~Sco. Here we show that the complex and previously-unexplained photometric behavior of V1309~Sco prior to its main outburst results naturally from the runaway loss of mass and angular momentum from the outer Lagrange point, which lasts for thousands of orbits prior to the final dynamical coalescence, much longer than predicted by contemporary models. This process enshrouds the binary in a ``death spiral'' outflow, which affects the amplitude and phase modulation of its light curve, and contributes to driving the system together. The total amount of mass lost during this gradual phase ($\sim 0.05\,\msun$) rivals the mass lost during the subsequent dynamical interaction phase, which has been the main focus of ``common envelope'' modeling so far. Analogous features in related transients suggest that this behavior is ubiquitous.
	\end{abstract}

\section{Introduction}

During some point in their lives, more than a quarter of massive stars, and a comparable fraction of their lower mass brethren, undergo direct interaction with a companion binary star, in a process commonly known as ``common envelope'' evolution \citep{sana12,kochanek14}.  Though short-lived and poorly understood, the outcome of this phase has crucial implications for all stages of stellar evolution.  The binary components may survive as distinct entities with a reduced orbital separation, leading in some cases to compact object binaries composed of white dwarfs, neutron stars and black holes.  A fraction of these systems will eventually merge through gravitational wave emission, as observed recently with the LIGO and Virgo interferometers \citep{ligoa,ligob,ligoc}. Alternatively, the stellar binary can coalesce completely into a single object with potentially exotic properties. The physical processes at work during the merger which delineate these two outcomes remain one of the biggest unsolved problems of stellar evolution \citep[e.g.][]{paczynski,webbink84,livio88,ivanova}.  Mass ejection during the terminal phase of these events can power bright transient emission \citep{ivanovasci,soker06}, offering a direct observational window into some of these open issues.

Catastrophic binary interactions have been suggested as an underlying driver for a class of optical transients commonly known as Luminous Red Novae (LRN) or Intermediate Luminosity Optical Transients (ILOT). This modest but growing group includes V838~Mon, V4332~Sgr, OGLE-2002-BLG-360, M31 LRN 2015, M101 OT2015-1, and NGC 4490 OT2011 \citep[e.g.][]{macleod17,mauerhan17,smith16,soker03,tylenda06,tylenda13,kurtenkov15,williams15}. The case for binary interaction as the origin of LRN/ILOT was significantly strengthened with the discovery of V1309~Sco, where the archival OGLE photometry revealed a contact binary with a decaying orbital period that was followed by a luminous outburst \citep{tylenda11}. Below we review the pre-outburst evolution of V1309~Sco and outline the structure of this paper.

\subsection{Overview of V1309 Sco light curve}

\begin{figure*}
\centering
\includegraphics[width=0.9\textwidth]{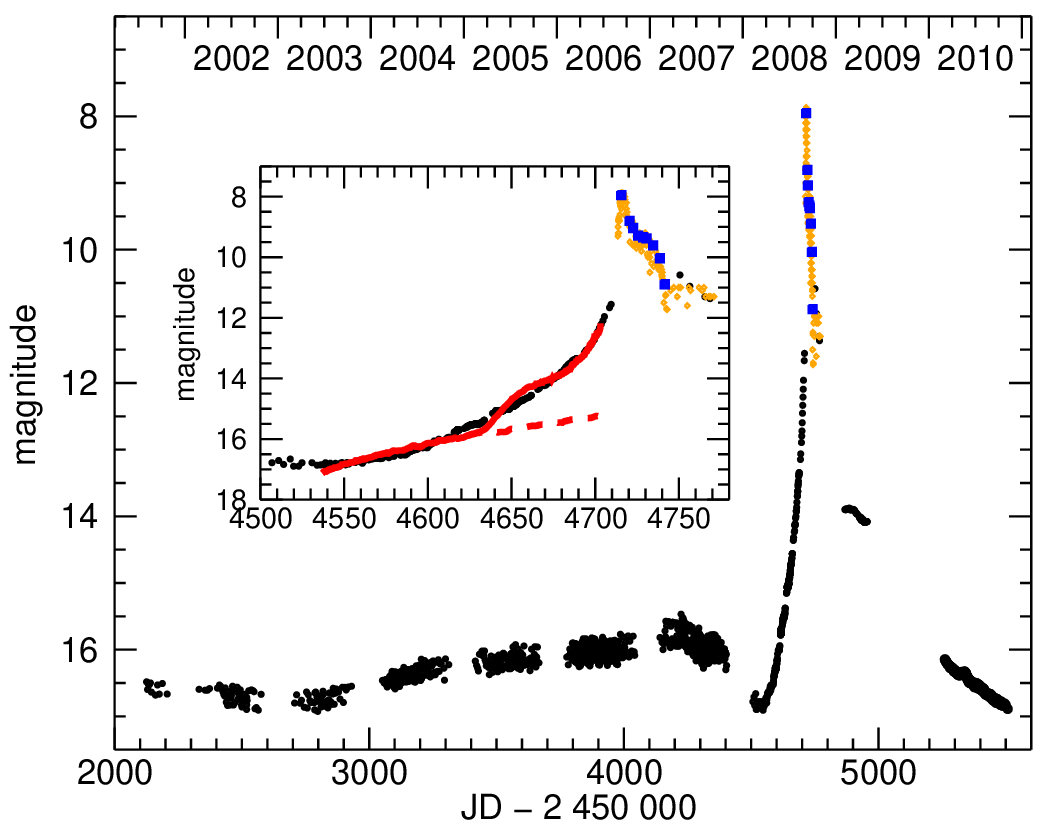}
\caption{Light curve of V1309~Sco. Black points show $I$-band observations from OGLE \citep{tylenda11}; blue squares show $V$-band magnitudes from ASAS\cite{pojmanski97}; and orange points show visual magnitudes from AAVSO \citep{kafka17}. The inset plot shows the observed gradual rise of brightness compared to bolometric light curves $-2.5 \log L_{\rm bol}$ from our three-dimensional SPH simulations, both with (solid red) and without (dashed red) accounting for evolution of the particle injection temperature. The simulated light curves are offset by an arbitrary constant to roughly match the observed magnitude at the onset of the slow brightening. The luminosity from the simulation ($20\,\lsun$) roughly matches the inferred luminosity of V1309~Sco binary ($10\,\lsun$) around the epoch JD 2 454 550 \citep{tylenda11,tylenda16}.}
\label{fig:lc}
\end{figure*}

\begin{figure*}
\centering
\includegraphics[width=0.9\textwidth]{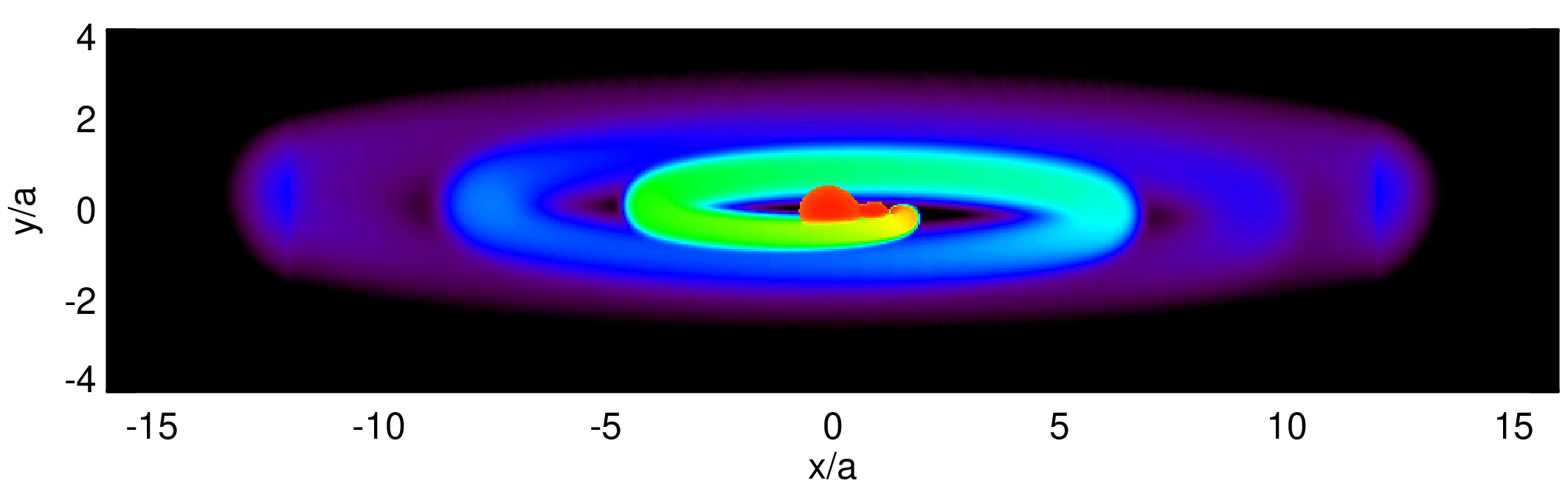}
\includegraphics[width=0.48\textwidth]{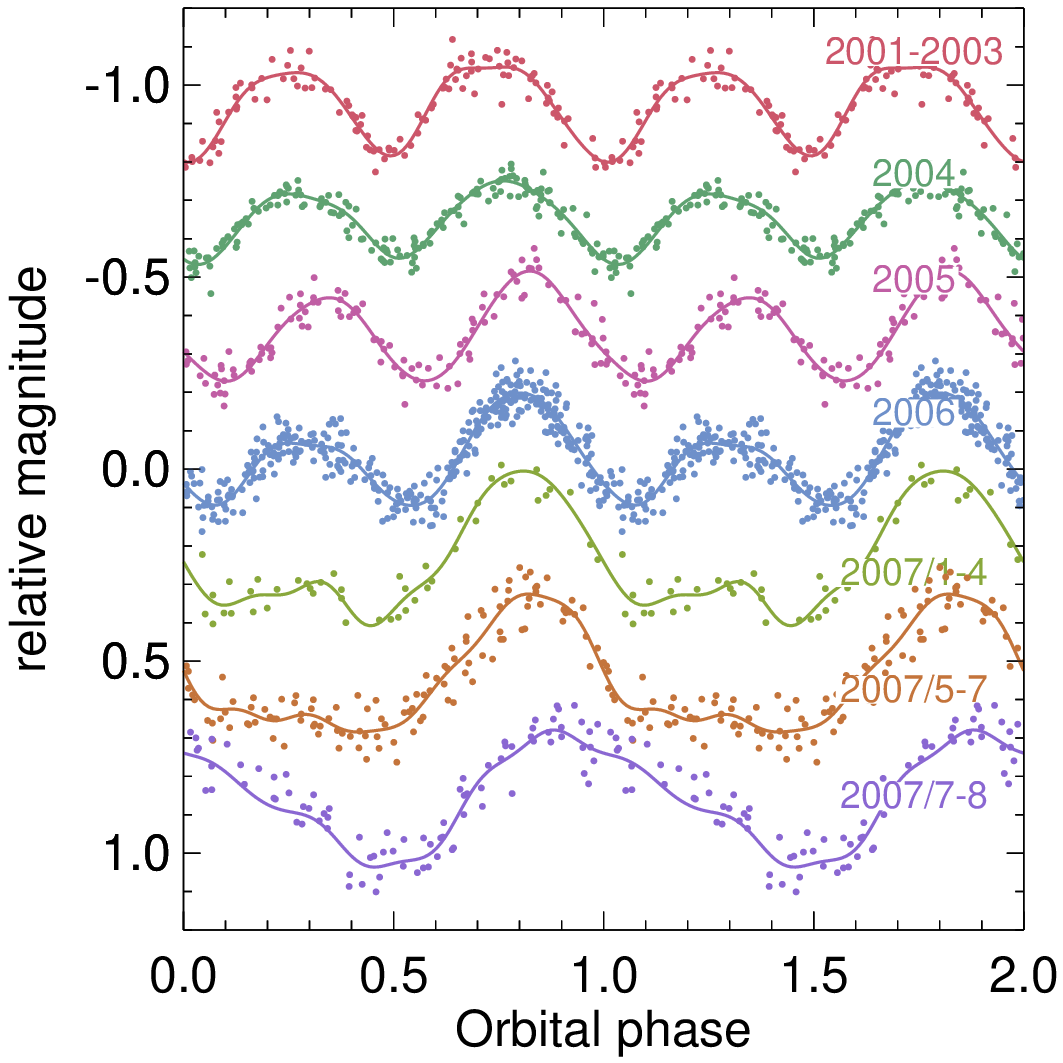}
\includegraphics[width=0.48\textwidth]{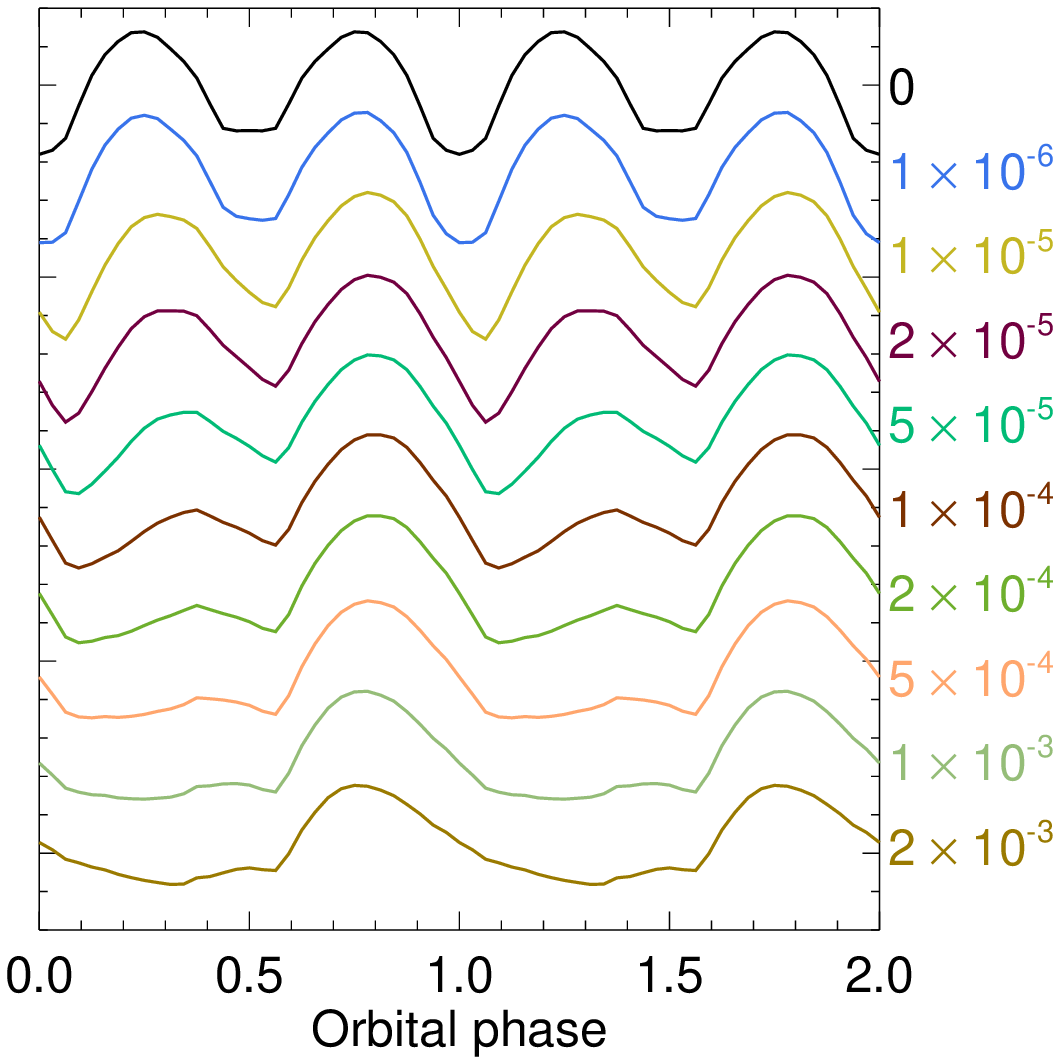}
\caption{Phased light curves of a V1309 Sco. ({\em Lower left}) Observed evolution from OGLE phased according to the time-changing orbital period \citep{tylenda11}. ({\em Lower right}) Theoretical light curves accounting for obscuration by the \ltwo stream, constructed for a binary of mass $M=1.65\,\msun$, mass ratio $q=0.1$, and semi-major axis $a=0.03$\,AU viewed at an inclination of $i=84^\circ$. Colors distinguish different values of the mass-loss rate $\mdot$, labeled in units of $\myr$. Light curves are offset vertically for purposes of clarity. ({\em Top}) Example ray-traced image showing the binary partially obscured by the \ltwo stream. An animated version of this figure is available in the on-line material.}
\label{fig:phase}
\end{figure*}

V1309~Sco was discovered near the peak of its outburst in September 2008 (Fig.~\ref{fig:lc}) and was soon spectroscopically recognized as a member of LRN \citep{mason10}.  Archival observations of V1309~Sco, primarily from the OGLE survey, revealed a contact eclipsing binary with an orbital period $P\approx 1.4$\,days, total mass $M =1$--$2\,\msun$, effective temperature $4500$\,K and luminosity $L_{\rm binary} \approx 10\,\lsun$, as measured $\approx$ 7 years prior to the outburst \citep{tylenda11}.  Between 2001 and 2007, the orbital period of V1309~Sco decayed rapidly, with a timescale $P/\pdot$ that decreased from about 1000 to 170 years, with a total decrease in the period over this interval of $\sim 1\%$ \citep{pejchabury,tylenda11}. The measured acceleration timescale $\pdot/\ddot{P}$ of a few years greatly exceeded the orbital period of the binary, yet was much shorter than the timescale for tidally-induced orbital decay \citep{pejchabury}. Simultaneously with this period decay, the phased light curve gradually morphed from a double-hump profile (typical for contact binaries) to a single-hump shape (Fig.~\ref{fig:phase}) as the mean brightness increased by about 1\,mag (Fig.~\ref{fig:lc}). This transformation occurred gradually over thousands of orbits, during which time the binary separation shrank by only $\sim 1.5\%$.  Earlier work attributed this transformation to the appearance of hot spots on the binary surface \citep{tylenda11}; however, given freedom on the number of spots, as well as their sizes, temperatures, and time evolution, this phenomenological model can reproduce almost any light curve shape.  Below, we present an alternative explanation with fewer free parameters, which naturally explains other puzzling features of V1309~Sco.

During the second half of 2007, V1309~Sco began to fade and around the same time its orbital-timescale variability ceased.  Then, over the first 8 months of 2008, the system gradually brightened by $\gtrsim$ 5 mag, on a timescale which, while still much longer than the orbital period, was comparable to the period acceleration timescale $\pdot/\ddot{P}$.  Such a slow rise is perhaps unexpected: the initial rise time of ``explosive" astrophysical transients is usually fast, similar to the radial expansion timescale of the ejecta.   Finally, in early September 2008, V1309~Sco brightened by an additional $3$ magnitudes to its peak in just a few days.  The subsequent $\sim 20$-day long plateau was interpreted as emission produced by hydrogen recombination following the ejection of $0.04$--$0.09\,\msun$ from the binary \citep{ivanovasci,nandez}.

\subsection{Outline of the paper}

Although the physical mechanism responsible for driving V1309~Sco to merger was attributed to the tidal Darwin instability \citep{stepien11,nandez,rasio95}, this alone cannot account for the complex evolution of the binary orbital period, phased light curve, and mean brightness.  It is also unclear how the final dynamical event was triggered: what, precisely, was so special about the point at which the binary evolution became dynamical, given that preceding changes to the orbit and structure occurred gradually of thousands of orbits?  We will argue that mass-loss from the outer Lagrange point of the binary \ltwo likely plays the dominant role in this behavior.  \ltwo mass loss was previously invoked to explain the pre-maximum behavior of V1309~Sco \citep{tylenda11,tylenda16}, and was shown to work at the order-of-magnitude level \citep{pejchabury}.  

In Section~\ref{sec:ltwo} we review the physics of \ltwo mass loss and describe our radiation hydrodynamic simulations with realistic microphysics.  We show that by postulating mass loss occurred from \ltwo at a progressively increasing rate $\mdot$ leading up to the dynamical phase, one can account for the full sequence of behavior observed in V1309~Sco quantitatively, in detail, and with a smaller number of free parameters than previous models.  In Section~\ref{sec:phased} we show that the observed changes in the phased light curve due to obscuration by the mass loss streams are consistent with the same mass loss required to explain the period decay.  Then in Section~\ref{sec:brightening} we argue that the slow brightening toward maximum at later times is naturally explained as a continuation of the runaway behavior of $\mdot$.  Based on the results of our simulations, we speculate that what defined the final dynamical event was the critical point at which the binary had lost most of its angular momentum through \ltwo outflow.  Given that V1309~Sco represents the best studied member of its class, the ramifications from our results may extend to the broader class of transients resulting from stellar mergers, and more broadly to the theory of catastrophic binary interactions, as we discuss in Section~\ref{sec:discussion}.

\section{\ltwo mass loss}
\label{sec:ltwo}

\ltwo mass loss was originally introduced as a natural outcome of a wide range of binary evolution pathways \citep{flannery,webbink76,livio79}. However, because of the common belief that the two stellar components cannot maintain complete corotation with the orbit, frictional drag in a ``common envelope'' has been instead invoked as the main physical process responsible for tightening the binary orbit \citep{paczynski,meyer,ivanova,macleod15}. Some numerical hydrodynamical simulations find \ltwo mass loss, but only in the few final orbits before the dynamical interaction, especially in cases when the binary begins in synchronous rotation \citep{lombardi11,nandez,rasio95a}.  

We consider the following scenario to explain V1309~Sco. We suggest that V1309~Sco was initially a contact binary, which increased its fillout factor due to the primary evolving off of the  main sequence or due to shrinking of the orbit by the Darwin instability triggered by secular decrease of the mass ratio due to thermal relaxation oscillations. The structure of the binary was not markedly different from other field contact binaries, in particular the stars were in corotation with the orbit. This is consistent with the phased light curve in 2001--2003, relatively slow initial orbital period change consistent with tidal timescale, and the existence of other stable contact binaries with similar orbital period \citep[e.g.][]{pawlak13}. When the expansion of the binary surface reaches the outer critical surface, the binary starts losing mass from \ltwo\footnote{In principle, similar setting can occur in normal mass transfer through L$_1$, when the secondary cannot accept the mass and inflates, forming an object that presumably looks similar to a contact binary, as was considered for example by \citet{livio79}.}.  We postulate that the gas is launched at corotation from \ltwo and that the timescale of angular momentum loss is much longer than the orbital period. We note that, empirically, synchronous rotation appears to be an excellent assumption in V1309~Sco because of its nearly constant orbital period observed for several years prior to the merger by OGLE. Whether the mass lost through \ltwo comes from the lighter star or is first transferred from the heavier one to the lighter one is not clear, but this uncertainty is included in our discussion below.

In this Section, we briefly review the physics of \ltwo mass loss and describe our calculation of various quantities. We summarize our nomenclature in Table~\ref{tab}.

\subsection{Brief overview of previous results}

The physics of \ltwo mass loss is conceptually similar to the more commonly studied case of mass transfer through the inner Lagrange point.  Considering the ballistic motion of cold matter, the spiral stream produced by matter leaving in co-rotation from \ltwo experiences tidal torques from the binary and gains sufficient energy to become unbound for binary mass ratios $0.064\le q \le 0.78$, reaching asymptotic velocities up to $\sim$25\% of the binary escape velocity.  Mass loss from binaries with mass ratios outside of this range instead was predicted to form a bound circumbinary disk \citep{shu79}.  We recently showed using three-dimensional smoothed-particle hydrodynamic simulations including the effects of radiative cooling and diffusion that \ltwo mass-loss can lead to qualitatively different outcomes, depending on the stellar surface temperature (initial temperature of matter at \ltwo) and cooling efficiency, in addition to the binary mass ratio \citep{pejchamb}.  In cases which produce unbound streams, internal shocks between the merging spiral arms power a substantial luminosity $L_{\rm stream}$ in excess of that originating from the surface of the binary star \citep{pejchama}.

\subsection{Energetics of \ltwo mass loss}
\label{sec:energy}

Test particles positioned at \ltwo which are corotating with the binary always possess a negative sum of kinetic and gravitational potential energies.  When matter starts moving out on a ballistic trajectory, tidal torques from the time-changing gravitational field of the binary increase the total energy, such that the final asymptotic energy is positive for a wide range of binary mass ratios, $0.064 < q < 0.78$. This energy loss from the binary orbit $\dot{E}_{\rm \ltwo}$, mediated by gravitational field, can be expressed as
\begin{equation}
\frac{\dot{E}_{\rm \ltwo}}{E_{\rm orb}} = \left[-2(\mathcal{E}_f - \mathcal{E}_i)\frac{(1+q)^2}{q} \right] \frac{\mdot}{M},
\label{eq:energies}
\end{equation}
where $E_{\rm orb} = -GM_1 M_2/(2a)$ is the energy of the orbit, and $\mathcal{E}_f - \mathcal{E}_i$ is the difference between initial and final energy of test particle in units of $GM/a$;  $\mathcal{E}_i < 0$ for all $q$ in corotation, $\mathcal{E}_f > 0$ for $0.064 < q <0.78$ \citep{shu79}.  Using previous analytic calculations \citep{shu79}, the term in square brackets has a numerical value of $-4.48$ for $q=0.1$.

At the same time, the loss of angular momentum from \ltwo and the requirement of circular orbit imply that the orbital period will decrease according as
\begin{equation}
\frac{\pdot}{P} = f\mathcal{A}\frac{\mdot}{M},
\label{eq:pdot}
\end{equation}
where $\mathcal{A}$ is a constant reflecting the specific angular momentum of the \ltwo point, which depends on $q$ and which of the two stars is actually losing mass \citep{pribulla98}, and $f$ is a factor representing the additional angular momentum extracted from the binary by tidal torques. We choose a value of $f=1.2$ based on our previous results \citep{pejchama}, although this is close to the maximum allowed value \citep{shu79}. Together with the third Kepler law, this angular momentum loss implies a corresponding orbital energy loss
\begin{equation}
\frac{\dot{E}_{\rm orb}}{E_{\rm orb}} = \left[\frac{1+q}{q}\frac{\mdot_1}{M} + (1+q)\frac{\mdot_2}{M} - \frac{2}{3}f\mathcal{A} - \frac{1}{3}  \right]  \frac{\mdot}{M}.
\end{equation}
For $q=0.1$ and a fiducial value of $f=1.2$, we find that the term in square brackets takes on values of $-9.6$ and $-43$ for cases when the total mass loss occurs entirely from the lighter ($\mdot_1=\mdot$) or heavier star ($\mdot_2=\mdot$), respectively. Even for $f=1$, the coefficient values are $-6.2$ or $-36$, respectively.

These results imply that the loss of binary orbital energy is more than enough to unbind material through \ltwo mass loss. Most of the energy gain  $\mathcal{E}_f - \mathcal{E}_i$ experienced by gas flowing out of \ltwo is used on countering the gravitational potential well. The asymptotic kinetic energy is at most $\sim 30\%$ of the total energy gain  $\mathcal{E}_f - \mathcal{E}_i$. However, for $q=0.1$ it is only $7\%$ \citep{shu79}.

The remaining released orbital energy, which was not expended on unbinding and accelerating the outflowing mass by tidal torques, presumably goes to heating and expansion of the stars, as well as the kinetic energy of bulk motions in the stellar surface layers. Additional work is needed to fully understand the partitioning of the orbital energy, which ultimately powers the envelope mass ejection.

\subsection{Smoothed-particle hydrodynamics of \ltwo mass loss}
\label{sec:sph}

Our numerical simulations of \ltwo mass loss were conducted with a custom smoothed-particle hydrodynamics (SPH) code developed to study dynamics and radiative properties of outflows from binary stars \citep{pejchama,pejchamb}. The problem is set up by continuously injecting particles with a specified temperature $T_{\rm in}$ in a small vicinity of the \ltwo point with an initially Gaussian density profile perpendicular to the binary axis. The initial Gaussian width $\varepsilon=c_S/v_{\rm orb}$ is roughly set by the ratio of the stellar surface sound speed $c_S$ and orbital velocity $v_{\rm orb}$ \citep{shu79}, and we have verified our results are insensitive to the value of $\varepsilon$ as long as it is sufficiently small. We use $\varepsilon = 0.01$ as our fiducial value. Particles which cross back inside of \ltwo are removed from the simulation domain; particles are adaptively injected to ensure that total active mass in the simulation corresponds to total mass lost from the binary. In the case of a time-dependent $\mdot$, we vary the mass of injected particles to maintain specified particle injection rate, which sets the numerical resolution of the simulation. Since we primarily study unbound \ltwo outflows, all particles within a smoothing volume were typically ejected around the same time and have similar mass.

The particles are initially in corotation with the \ltwo point and their acceleration is calculated using standard hydrodynamical and viscous forces with adaptive smoothing length.  Gravitational forces from the binary star are modeled as a combination of two point masses, $M_1$ and $M_2$, with total mass $M = M_1+M_2$, mass ratio $q = M_1/M_2$, semi-major axis $a$ and orbital period $P$.  Self-gravity between the particles is neglected\footnote{A simple comparison between the timescale for gravitational collapse to that of outflow expansion suggests that the gas will not form gravitationally-bound substructure \citep{pejchama}. Thermal instability is a more promising way to form clumps in the outflow, which together with associated changes in the opacity provides a possible explanation for the small-scale ``wiggles'' that occur during the rising phase of the light curve \citep{pejchabury,pejchama}.}.  The energy evolution includes hydrodynamic and viscous terms, flux-limited radiative diffusion, and radiative cooling \citep{forgan09,stamatellos07}. Calculations relevant for the phased light curve evolution include also irradiation by the central binary, which we model as three point sources positioned at the barycenter, and slightly above and below to better capture vertical extent of the source. The implementation of these effects, equation of state, opacities, and estimates of luminosity and effective temperature are mostly identical to our previously published works \citep{pejchama,pejchamb}.  Even our approximate implementation of the relevant radiative processes results in an outflow temperature structure, which is coupled to the radiation and is therefore significantly more realistic than what would be obtained from a purely adiabatic treatment. This allows us to obtain phased light curves by post-processing the simulation data, as detailed below.

\subsection{Calculation of phased light curves}

In order to build our understanding of how the phased light curve of the binary changes when viewed through the \ltwo stream, we constructed a semi-analytic model of the problem. We calculate the trajectory of the stream center, as well as the velocity $v$ and the evolution of perpendicular stream spread $\epsilon$ \citep{shu79}, which therefore gives also the density at the stream center 
\begin{equation}
\rho_{\rm center} = \frac{\mdot}{2\pi \epsilon^2 v}.
\label{eq:rho_center}
\end{equation}
The density profile moving away from the stream center is assumed to obey 
\begin{equation}
\rho(\psi) = \rho_{\rm center} \exp\left(-\frac{\psi^2}{2\epsilon^2}\right),
\label{eq:rho_psi}
\end{equation}
where $\psi$ is the distance from a given position to the closest point on the stream center trajectory. The temperature along the stream center is set by irradiation from the binary
\begin{equation}
T = T_{\rm binary}\sqrt{\frac{r_{\rm \ltwo}}{r}},
\label{eq:t_binary}
\end{equation}
where $r_{\rm \ltwo}$ is the distance of the \ltwo point from the barycenter; a fixed temperature is assumed for all values of $\psi$. The stream is truncated at the radius $r_{\rm coll}$ where consecutive spiral arms collide, which is one of the free parameters of the model with a default value of $8a$ as motivated by the collision radius seen in our simulations \citep{pejchama}. The density profile outside of $r_{\rm coll}$ is assumed to be an axially-symmetric wind with constant $\mdot$ ($\rho \sim r^{-2}$) and opening angle set to match the value of $\epsilon$ at the radius of the collision.  We neglect contributions to the luminosity which arise from the collision of the  spiral streams, because at times of interest we find that the collision luminosity is at most comparable to the binary luminosity; furthermore, the low value of $\teff \approx 2000-3000$\,K of this emission (due to the large collision radius) will suppress its contribution to the optical bands for $\mdot$ achieved before 2008.  With these informed choices, we find a  remarkably accurate match to the density and temperature profile of matter surrounding the binary compared to more complete SPH calculations.

The appearance of the  binary surrounded by the \ltwo mass loss stream is constructed by semi-implicitly integrating the intensity $I$ along parallel rays according to the standard radiative transfer equation
\begin{equation}
I(s+ds) = e^{-\kappa\rho ds}I(s) + (1-e^{-\kappa\rho ds})S(s),
\label{eq:ray}
\end{equation}
where $s$ is measured along the ray and $S(s)$ is the source function.  The latter we assume to either obey $S \propto T^4$ when obtaining the bolometric light curve, or the Planck law $S = B_\lambda(T)$ when considering the emission at specific wavelengths. The results shown for V1309~Sco are calculated for $I$ band at $\lambda = 800$\,nm, corresponding to the OGLE observations.  We assume that the opacity of gas with hydrogen mass fraction $X=0.7$ and metallicity $Z=0.02$ is given by the approximate expression
\begin{equation}
\kappa = \kappa_{\rm m} + \left(\kappa_{H^-}^{-1} + (\kappa_{\rm e} +\kappa_{\rm K})^{-1}\right)^{-1},
\end{equation}
where $\kappa_{\rm m} = 0.1Z$ is an approximate molecular opacity, $\kappa_{H^-} = 1.1\times 10^{-25} Z^{0.5} \rho^{0.5}T^{7.7}$ is the $H^-$ opacity, $\kappa_{\rm e} = 0.2(1+X)$ is the electron scattering opacity, and $\kappa_{\rm K} = 4\times 10^{25} (1+X)Z \rho T^{-3.5}$ is the Kramers opacity, all in cgs units. The default opacity for $r>r_{\rm coll}$ is $\kappa_{\rm wind} = \kappa_{\rm m}$, but we investigate both larger and smaller values in order to study, in a heuristic manner, effects such as dust formation.  

The rays are cast on a regular grid, the orientation of which we vary relative to the binary orbital plane to study the appearance of the system from different inclinations $i$ and phase angles $\varphi$. If a ray intersects the outer critical surface of the binary, we initiate the integration of Equation~(\ref{eq:ray}) at this point with initial temperature $4500$\,K and take into account linear limb darkening law, where we use the angle between the ray and the normal to the outer critical surface. If a ray does not intersect the binary surface, we put the starting point on the far side of the binary with intensity initially set to zero.

\subsection{Calculation of gradual brightening}
\label{sec:calc_brightening}

When directly simulating the final gradual rise toward the maximum, the parameters of the binary will significantly change due to the loss of mass and angular momentum. We prescribe the time evolution $\mdot(t)$ as a power-law approach to a singularity \citep{webbink77b,pejchabury},
\begin{equation}
\mdot(t) = \mdot_0 \left(\frac{t_0}{t_0 - t}  \right)^\delta.
\label{eq:mdot}
\end{equation}
As discussed in detail below, we find that in order to match the light curve we need to eventually increase the injection temperature of the particles $T_{\rm in}$. We assume that $T_{\rm in}(t)$ evolves as
\begin{equation}
T_{\rm in}(t) =
\begin{cases}
T_{{\rm in},0} &  \mdot(t) < \mdot_T \\
T_{{\rm in},0} \left(\frac{\mdot(t)}{\mdot_T}  \right)^\gamma  & \mdot(t) \ge \mdot_T
\end{cases}
\label{eq:tin}
\end{equation}
We vary the parameters $\mdot_0$, $t_0$, $\delta$, $\mdot_T$, and $\gamma$ to match the observed light curve. $T_{{\rm in},0}$ is set to 4500\,K. 

The loss of angular momentum is self-consistently included using the rate of \ltwo mass loss according to Equation~(\ref{eq:pdot}). We experimented with several choices for the value of $\mathcal{A}$, corresponding to the assumption that one of the stars carried all of $\mdot$; $\mdot$ was split equally between the stars; or the mass loss was such that $q$ remains constant.  Each possibility gave similar results to within a factor of $\lesssim 2$. In the end, we choose to assign all of the mass loss to the lighter star, which produces the smallest change of orbital period for a given value of $\mdot$ \citep{pribulla98}. Although the mean density of the lighter star is higher, it might be very centrally condensated with small density in the envelope \citep{paczynski07}. More work is needed to understand which star is ultimately losing mass in this case. The orbital phase $\varphi$ of the binary is simultaneously integrated as $\dot{\varphi} = 1/P$. The luminosity produced by the outflow is a sum of radiative cooling rate of all particles.

\section{Changes in the phased light curve}
\label{sec:phased}

In this Section, we describe how runaway \ltwo mass loss explains the evolution of the phased light curve of V1309~Sco.

\subsection{Results}
\label{sec:phased_results}

Figure~\ref{fig:phase} shows how the phased light curve of the binary changes when viewed through the \ltwo mass loss stream, a ``death spiral'' with nearly edge-on inclination. In isolated binaries, the transit of the  smaller, less massive secondary in front of the larger, more massive primary is observed as the primary eclipse (phase $0$). After this point, the visible surface area of the binary increases and reaches a maximum at phase $0.25$. However, in this case the \ltwo stream trails behind the secondary and obscures the binary light, leading to a suppression of the maximum at phase $0.25$.  Half a period later (phase $0.75$), when the binary surface area again reaches maximum, the \ltwo stream is less opaque and positioned farther away from the binary, such that the observed flux at this phase is closer to its original, unattenuated value. The strength of this asymmetric effect increases as $\mdot$ rises in time, naturally explaining the transformation of the V1309~Sco light curve from double- to single-humped.

\begin{figure}
\includegraphics[width=0.45\textwidth]{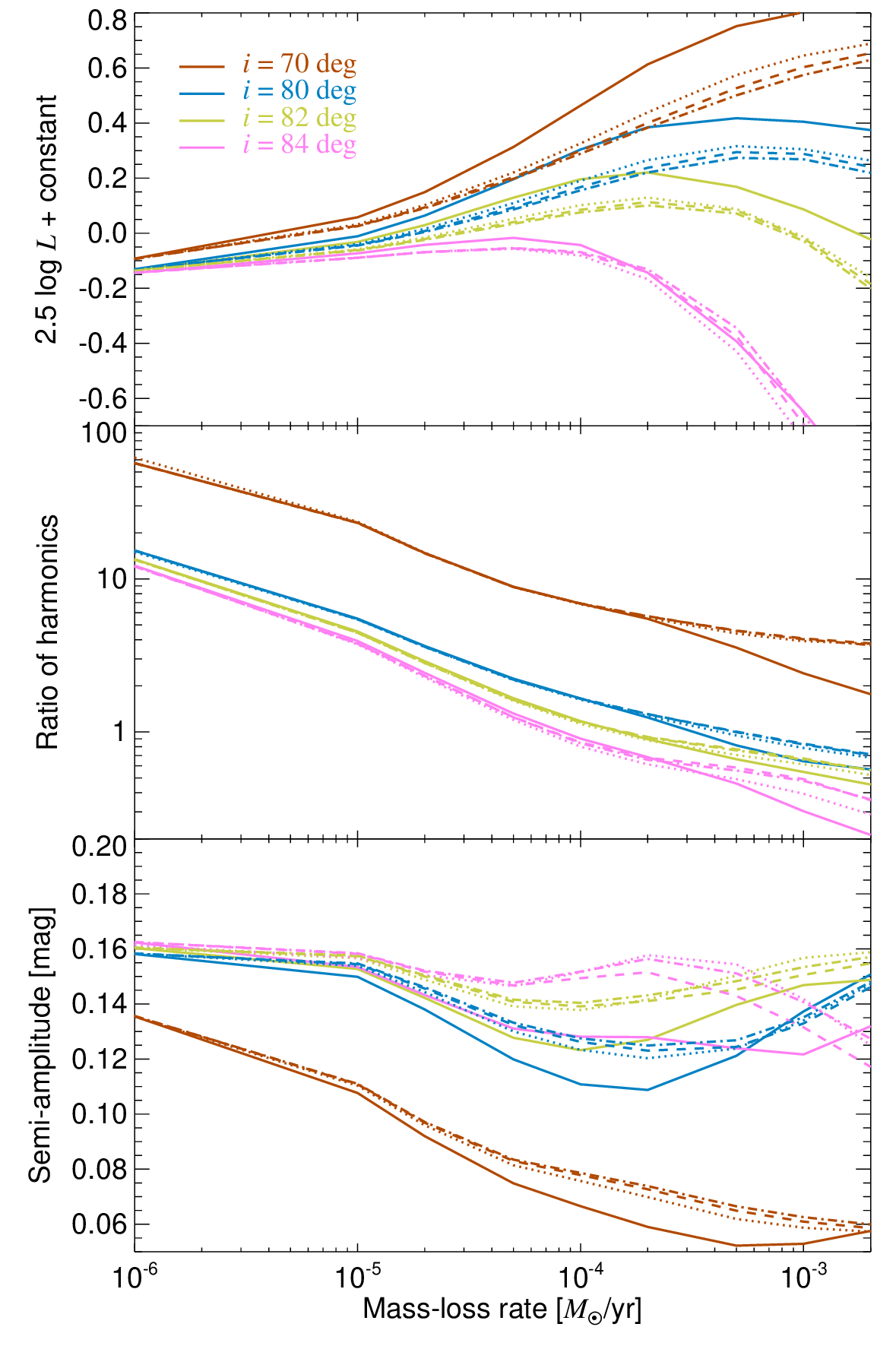}
\caption{Changes of observable quantities of phased light curves within our semi-analytic model as a function of $\mdot$: mean flux evaluated at $800$\,nm  (top), ratio of amplitudes of the second and first harmonics of the Fourier decomposition of the  light curves (middle), and the semi-amplitude in magnitudes (bottom). Colors indicate range of inclination angles as shown in the legend in the top panel and line styles denote different values of $r_{\rm coll}$ of $6a$ (solid), $8a$ (dotted), $10a$ (dashed), and $12a$ (dot-dashed).}
\label{fig:lc_anal}
\end{figure}

We illustrate this transformation of the phased light curve quantitatively in the middle panel of Figure~\ref{fig:lc_anal} for a range of binary inclinations and $r_{\rm coll}$. We first decomposed the theoretical light curves in harmonic sine and cosine waves with amplitudes $(b_j, c_j)$ and then calculated the ratio of second to first harmonic $\sqrt{(b_2^2 + c_2^2)/(b_1^2+c_1^2)}$. For contact binary with two maxima per period, the second harmonic dominates leading to very high values of the parameter. As the asymmetry due to obscuration by the \ltwo stream becomes stronger, the first harmonic comes to dominate, as seen in Figure~\ref{fig:lc_anal}.  At the same time, the light curve amplitude changes only slightly for greater inclination angles, for which the obscuring effect dominates.

The stream intersects the sightline to the binary for inclination angles $i \gtrsim 70^\circ$, and results in complete obscuration for $i \gtrsim 85^\circ$ (see also Sec.~\ref{sec:degeneracies}).  The lower limit on $i$ required to explain the light curve phase evolution breaks the degeneracy between the inclination and mass ratio in light curves of contact binaries \citep{rucinski93,rucinski01}, constraining the mass ratio of V1309~Sco to be $q \lesssim 0.1$ given the observed amplitude of the light curve variation in 2001. A lower limit on the mass ratio of $q \gtrsim 0.07$ can also be placed by requiring that the \ltwo spiral stream does not form a bound circumbinary disk, which instead would result in a much higher radiative efficiency than observed.

\begin{figure}
\includegraphics[width=0.45\textwidth]{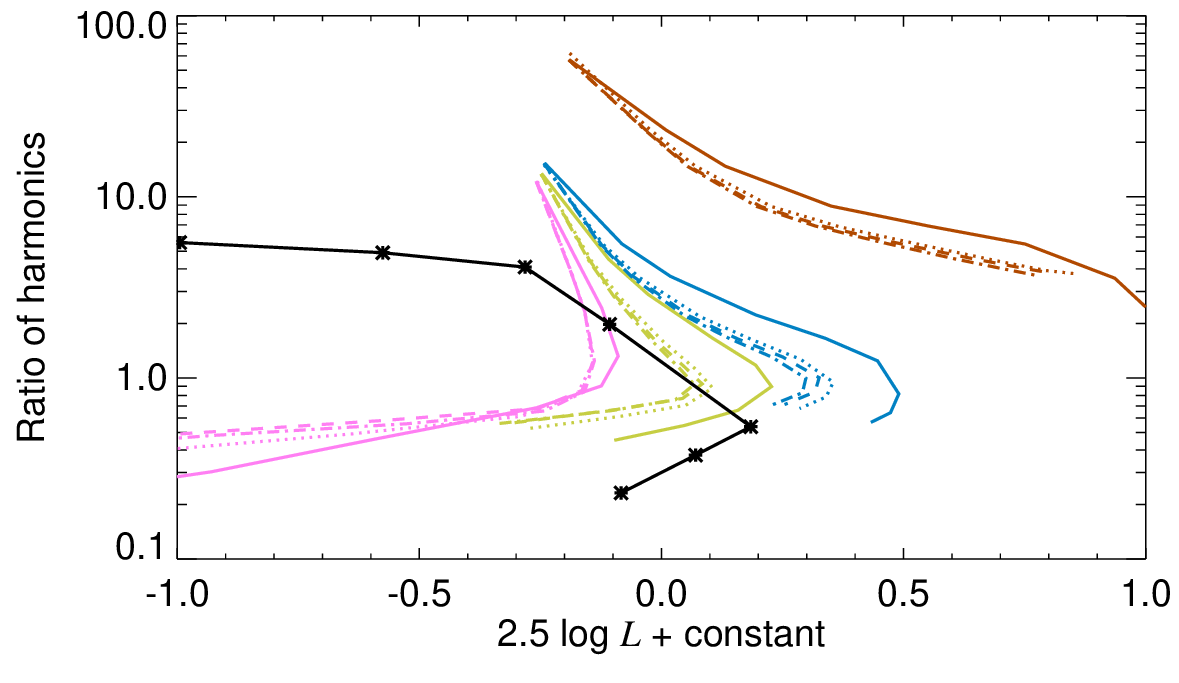}
\caption{Correlation between the mean flux and the ratio of amplitudes of second and first harmonics from Figure~\ref{fig:lc_anal}. The black points connected by lines indicate the same quantities obtained from the 2001--2007 OGLE light curves of V1309~Sco. To obtain these points, we first fitted the light curve with a spline to determine long-term flux variations. Next, we subtracted these variations and analyzed the phased light curves. For each line, the evolution in time proceeds from top to bottom.}
\label{fig:lc_anal_corr}
\end{figure}

As $\mdot$ grows, the additional contribution from radiative cooling of the \ltwo stream initially increases the total luminosity of the system.  However, this trend temporarily reverses once the central binary becomes almost completely obscured, after which point an even larger value of $\mdot$ further attenuates the luminosity. We illustrate this quantitatively in the top panel of Figure~\ref{fig:lc_anal}. Similar gradual brightening followed by subsequent dimming was observed in V1309~Sco in 2001--2007, although the total amplitude of this effect in our models is smaller than that observed.  A detailed comparison between the theory and the data on V1309~Sco is shown in Figure~\ref{fig:lc_anal_corr}. Part of the discrepancy might be due to changes in the surface temperature and luminosity of the binary as it undergoes rapid mass stripping, a possibility we investigate in Section~\ref{sec:response}.

\subsection{Parameter dependence and degeneracies}
\label{sec:degeneracies}

Except for small systematic offsets, the general evolution of the light curve, in particular its transformation from a double- to single-humped profile shown in Figure~\ref{fig:phase}, does not depend on the stream collision radius $r_{\rm coll}$ (Fig.~\ref{fig:lc_anal}). 
The secular increase in the mean observed flux depends on the assumed stream cooling efficiency.  The subsequent dip in the flux depends on the wind opacity $\kappa_{\rm wind}$, but is present even in the limit that $\kappa_{\rm wind}=0$. 

Thus far we have neglected the presence of dust along the path of the outgoing light from the binary. However, because the dust opacity $\kappa_{\rm dust} \gg \kappa_{\rm m}$, even a moderate amount of dust formation would completely obscure the central area.  A simple estimate, based on requisite conditions for dust condensation and coagulation, suggests that dust formation should be copious in the outflow given its density and expansion rate predicted by our model \citep{pejchama,pejchamb}. However, the process of dust formation in this setting is particularly complex and is controlled by the interplay between rapid expansion, shocks from spiral collisions, and irradiation by the central binary, all of which are treated primitively in our model.  Consistency with the observations in fact only requires that dust formation be suppressed in the irradiated ``skin" of the outflow, at radii within a few tens of $a$ (dust can still form in the orbital plane, where it would not directly obscure the central regions).  Such a picture agrees with modeling of the optical to infrared spectral energy distribution in 2007, for which the optical depth to the star was inferred to be $<1$, but a better fit to the observations is obtained if the material is arranged in a disk-like structure with a small vertical aspect ratio \citep{tylenda16}.  Regardless of these issues, dust formation probably cannot be avoided once the stream self-obscures and the binary variability vanishes ($\mdot \gtrsim 10^{-3}\,\myr$ or after JD 2 454 500 in V1309~Sco).  After this point, we are anyways primarily observing light which is generated by the spiral stream collisions well outside of the central star.

A degeneracy exists between three quantities which control the light curve appearance: $\mdot$, $i$, and $\kappa$. Our assumed opacity roughly corresponds to the Rosseland mean, but this should be regarded as a lower limit, because for values of $\mdot$ of interest the stream is not fully optically-thick. It is then easy to imagine that more appropriate band-integrated opacity would be a factor of a few times higher than the Rosseland mean \citep{alexander94}. As a result, the value of $\mdot$ that we infer from the phased light curve may be somewhat overestimated. On the other hand, a higher value of $\mdot$ would move the isodensity contours further from the orbital plane, such that the binary light curve will be noticeably affected even for smaller values of $i$ than needed in our fiducial model. Here, reasonable values of $\mdot$ imply we are viewing V1309~Sco at $80^\circ \lesssim i \lesssim 85^\circ$, but this depends sensitively on what we assume for the density profile perpendicular to the stream trajectory. Without a detailed simulation of the stellar surface near \ltwo, one can only hypothesize the possible range of outcomes -- for instance, if the density profile decreases shallower than the Gaussian profile we have assumed, then the allowed range of $i$ will be greater. In all cases, however, the stream geometry requires a minimum value of $i \gtrsim 70^\circ$, such that similar behavior to that of V1309~Sco should be observed in up to $\sim 1/3$ of randomly-oriented binaries. As a result, some freedom exists in the value of $\mdot$ required to reproduce the phased light curves (Sec.~\ref{sec:mdot}).

We have also explored calculating phased light curves directly from our SPH simulations.  Although we obtain qualitatively similar results to the procedure described above, the simulations suffer from poor resolution away from the stream center and artifacts near the injection point.  Furthermore, the simulation must be run for a long time to overcome the initial transient (which lasts tens of orbits in some cases), because the explicit timestep is very short due to radiative diffusion and cooling \citep{pejchama}.

\subsection{Comparison with directly observed orbital period}
\label{sec:mdot}

\begin{figure*}
\centering
\includegraphics[width=0.8\textwidth]{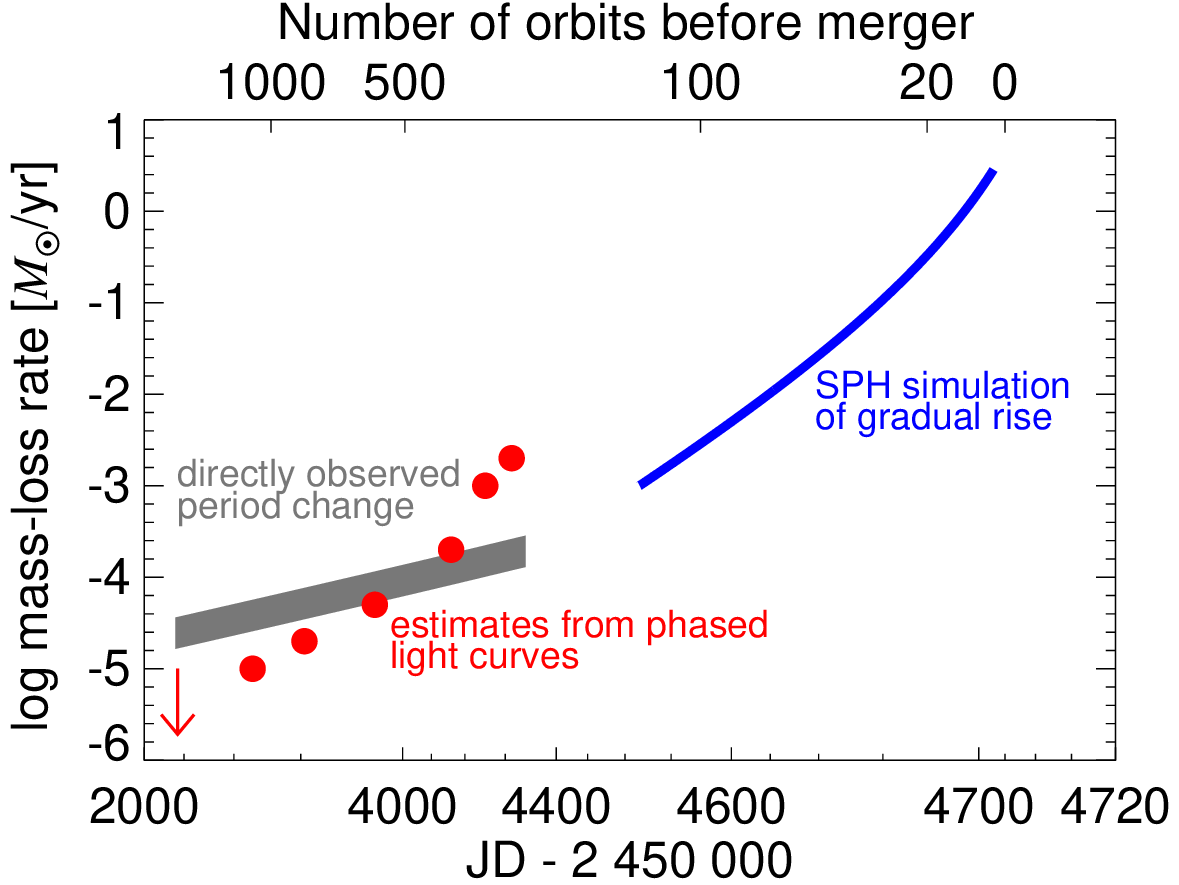}
\caption{Time evolution of the mass-loss rate in V1309 Sco derived from (1) the directly observed orbital period change assuming a binary mass ratio $0.07 \le q \le 0.1$ (gray line); (2) from visual comparison of theoretical and observed phased light curves in Fig.~\ref{fig:phase} (red circles), and (3) from simulations which reproduce the gradual slow luminosity rise to the maximum (blue line).  The approximate number of orbits remaining until the merger is indicated on the top axis. The width of the gray line includes uncertainties in assigning $\mdot$ to either star in the binary as well as those on the value of $q$. Sensitivity to the parameters and their degeneracy are discussed in Section~\ref{sec:degeneracies}. }
\label{fig:mdot}
\end{figure*}

The specific angular momentum extracted from the binary provides a direct link between the mass-loss rate $\mdot$ and the orbital period change $\pdot$ \citep{pejchabury,pribulla98} (Eq.~[\ref{eq:pdot}]). Figure~\ref{fig:mdot} compares the evolution of $\mdot(t)$ as inferred from the observed orbital period change to that required to reproduce the morphological changes observed simultaneously in the phased light curve. The good agreement between these two independent methods suggests that the orbital period change was driven primarily by \ltwo mass loss, at least at late times.  However, we cannot exclude the possibility that in the earliest phases (2001--2003) angular momentum loss was instead dominated by other processes, such as the Darwin tidal instability.

\section{Gradual rise to maximum}
\label{sec:brightening}

We hypothesize that nothing particularly dramatic occurred to V1309~Sco near the end of 2007: the system variability simply vanished once the binary surface became completely blocked by the \ltwo mass loss stream.  After this point, the observed emission was dominated by shocks internal to the \ltwo outflow instead of the binary surface.  Although near the end of 2007 the shock luminosity was comparable to that of the binary ($L_{\rm binary} \approx L_{\rm stream}$), the low temperature of the shock emission $\lesssim 3000$\,K (due to its comparatively large radius) resulted the $I$ band luminosity temporarily decreasing. The agreement between the predictions of our model and the observed binary evolution at low $\dot{M}$ motivates us to extend it to even higher mass loss rates. In this Section, we directly simulate the last $\sim 200$\,days of evolution with our SPH code assuming that the mass loss rate $\mdot(t)$ continues the runaway evolution inferred earlier, and then compare our results to the observed rising light curve prior to the dynamical phase.

\subsection{Results}
\label{sec:result_brightening}

Initially we experimented with varying only the parameters which describe $\mdot$ (Eq.~[\ref{eq:mdot}]) while keeping the matter ejection temperature $T_{\rm in}$ (Eq.~[\ref{eq:tin}]) fixed.  However, we found a limit to the amount of upward {\em curvature} which can be obtained in the light curve rise: no matter how large the value of $\mdot$ reaches, the escaping luminosity approaches a straight line, similar to the one shown with red dashed curve in the inset of Figure~\ref{fig:lc}. This is because, as $\mdot$ increases, the timescale for shock-powered radiation to diffuse out of the equatorial outflow also increases. Eventually, once the diffusion time exceeds the expansion timescale and the timescale over which $\mdot$ increase, this limits the rate at which the light curve can rise in the simulations.

\begin{figure*}
\centering
\includegraphics[width=0.8\textwidth]{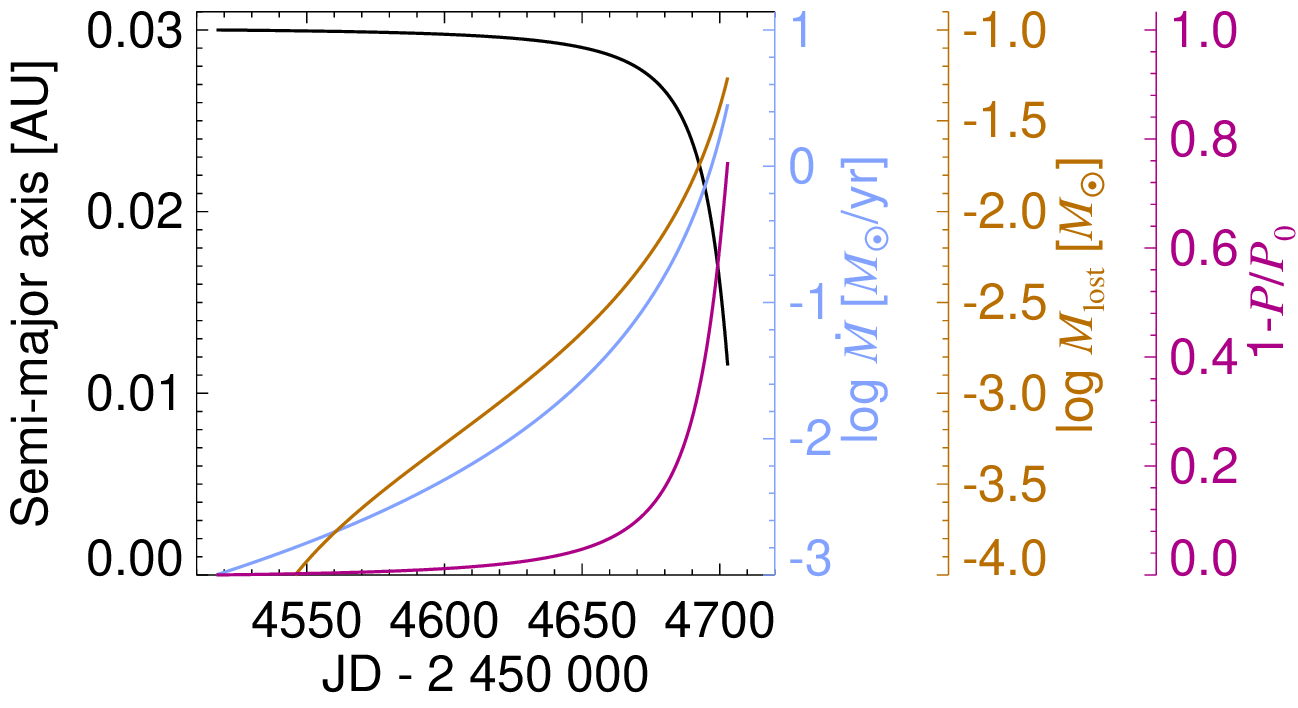}
\caption{Evolution of binary properties that reproduce the observed gradual accelerating rise seen in Figure~\ref{fig:lc}. Quantities shown include the semi-major axis $a$, mass-loss rate $\mdot$, total mass lost $M_{\rm lost}$, and the degree of asynchronism $1-P/P_0$.}
\label{fig:massloss}
\end{figure*}

\begin{figure}
\centering
\includegraphics[width=0.45\textwidth]{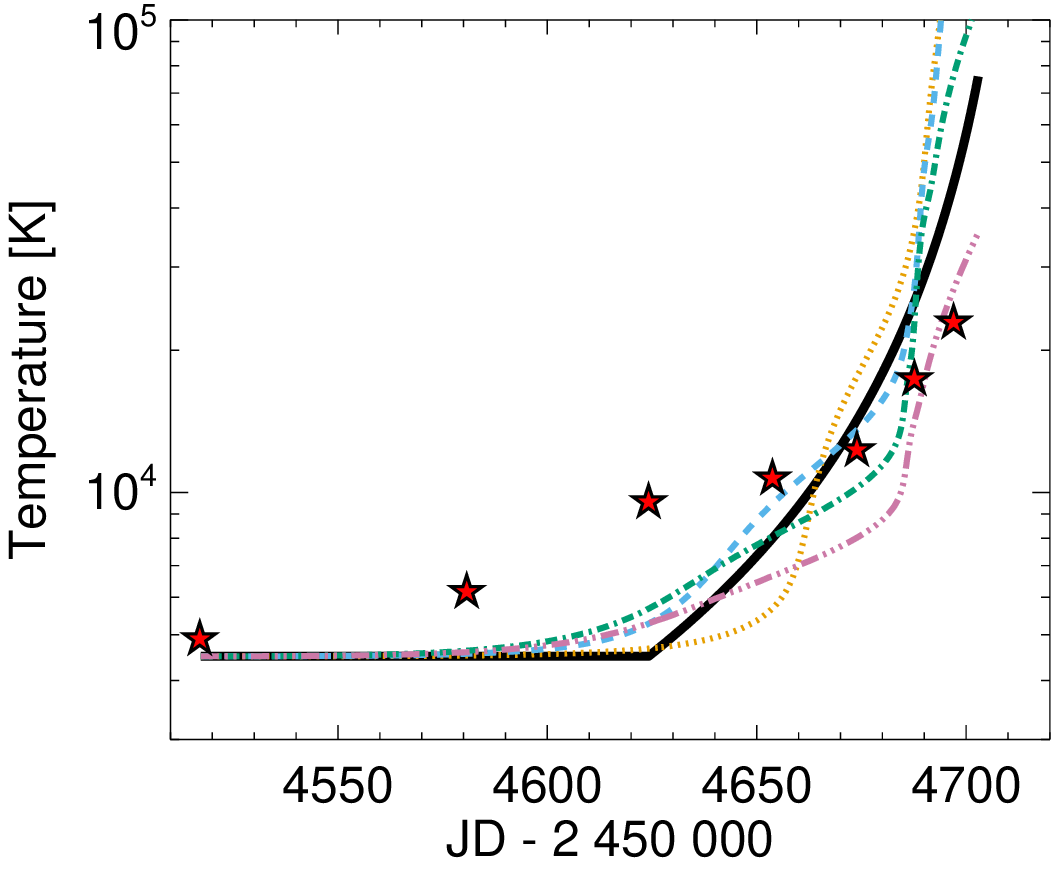}
\caption{Time evolution of the injection temperature of matter at \ltwo (black solid line) in comparison to our heuristic guess of its value due to frictional heating in the stellar envelope (colored lines).  The latter are determined by first using our equation of state to find internal energy corresponding to the binary surface temperature, for different assumptions of the density at which the energy is deposited: $\rho = 10^{-4}$ (orange dotted), $10^{-6}$ (blue dashed), $10^{-8}$ (green dot-dashed), and $10^{-10}$\,g\,cm$^{-3}$ (pink triple-dot-dashed). Although the surface density of a $10\,R_\odot$ subgiant from Section~\ref{sec:response} is within the assumed range, our choice of  density range but merely serves the point of illustrating that the results do not depend much on the density.   Then, adding the approximate rate of frictional energy deposition $\Delta v^2/2$, where $\Delta v = (1-P/P_0)v_{\rm orb,0}$ is the velocity shear due to desynchronization based on the observed evolution of V1309~Sco (Fig.~\ref{fig:massloss}), we interpolate back to find the new surface temperature. Wiggles in the lines result from the ionization of hydrogen and helium.  Stars show for comparison the predictions for the evolution of the effective temperature of a single non-rotating evolved star of mass $1 M_{\odot}$ and radius $10\,R_\odot$, which is subject to the inferred mass loss history shown in Fig.~\ref{fig:massloss}.}
\label{fig:tsurf}
\end{figure}

We find that the observed upward curvature of the light curve can be explained if we assume that the initial temperature $T_{\rm in}$ of the \ltwo mass loss also increases with $\dot{M}$, as shown with red solid line in the inset of Figure~\ref{fig:lc}.  We parameterize the temperature increase according to Equation~(\ref{eq:tin}), where we take values $\mdot_0 = 10^{-3}\,\myr$, singularity $t_0$ positioned $200$\,days after the start of the simulation, $\delta =3$, $\mdot_T = 10^{-2}\,\myr$, and $\gamma=0.5$. Not being the result of a rigorous fit, these parameters are probably degenerate to some extent, but limited computational resources precluded us from exploring the parameter space more completely. The temporal evolution of quantities of interest, such as $\mdot$, the total mass loss, and degree of binary asynchronism, are shown in Figures~\ref{fig:massloss} and \ref{fig:tsurf}. The overall mass-loss history prior to the main peak is  visualized in Figure~\ref{fig:mdot}. Our calculations suggest that V1309~Sco lost a total of $0.05\,\msun$ prior to the main peak. A movie showing the evolution of the \ltwo outflow structure in the final $\sim 200$\,days is available in the online materials.

We explored a number of alternative scenarios to explain the observed accelerating brightening before the main peak, but none were found to work. For instance, changes in the bolometric correction as the effective temperature of the equatorial outflow increases with growing $\mdot$ are insufficient to explain the upturn of the flux. As the binary becomes tighter, the velocity of the \ltwo increases in proportion to the orbital velocity; however, the total budget of angular momentum and mass in the binary does not allow the later faster ejecta to overtake the earlier slower material at radii where the collision could efficiently radiate the dissipated energy (a process which anyways would already be accounted for in our simulations).  Mass loss from the secondary could also reduce the binary mass ratio $q$ below the critical value of $\lesssim 0.064$ at which the \ltwo outflow is expected to transition from a comparatively radiatively-inefficient narrow equatorial outflow, to a more radiatively-efficient decretion disk or isotropic wind \citep{pejchamb}. However, the initial mass of the secondary would need to be fine-tuned to achieve this transition at just the right time to generate the required luminosity\footnote{Our final simulation shown in Fig.~\ref{fig:lc} evolves to $q<0.064$ only in the last day or so, which is too late to appreciably affect the radiative properties.}. Using dust-free opacity tables only shifts the problem to somewhat higher values of $\mdot$.

Our code eventually crashes and we cannot follow the evolution any further. Figure~\ref{fig:massloss} shows that this happens when the semi-major axis and orbital period have decreased to about a third and fifth of their original values, respectively.  Since the mass ratio does not change appreciably, the orbital angular momentum has also decreased to about the third of its original value by this point.  About $10$\,days after the termination of our simulation, V1309~Sco underwent its final dynamical phase and concomitant brightening to maximum.  It is thus tempting to speculate that the final dynamical interaction was a direct consequence of the nearly complete loss of the binary angular momentum by the prolonged \ltwo outflow.  However, without a direct simulation of the evolution of the binary structure itself we cannot definitively prove a causal link.

This speculation, however, gives an observably testable prediction: to remove the orbital angular momentum from the system, the binary needs to lose about $10$--$30\%$ of the mass of the lighter star by \ltwo outflow \citep{macleod17}.  Part of the uncertainty in the total mass lost comes from ambiguity in assigning the mass loss to either of the star in the binary \citep{pribulla98}. Furthermore, the mass loss rate culminates in the final 10--20 orbital revolutions, when the binary might not be able to maintain corotation with the orbit even in its surface layers. Mass lost during this phase would thus carry lower specific angular momentum than that of \ltwo in corotation, which would require a higher amount of mass to be lost. Nonetheless, many transients similar to V1309~Sco show evidence of secondary peaks which can be explained by shock interaction of dynamical ejecta with an equatorially-focused pre-existing material containing about 10\% of the binary mass \citep{metzger17}. In principle, more detailed modeling of the transient light curves and spectra can be used to constrain the quantity of pre-dynamical mass loss in the equatorial outflow.

\subsection{Interpretation of the increase of $T_{\rm in}$}

As described in Section~\ref{sec:result_brightening}, the most plausible scenario we find to explain the accelerating brightening of V1309~Sco in early 2008 is that, in addition to the rising value of $\mdot$, the temperature $T_{\rm in}$ at which matter is ejected at \ltwo must increase secularly as well.  As a result of the rising temperature, the spiral stream becomes progressively wider, the self-collision occurs closer to the binary, and the radiative efficiency $L_{\rm stream}/\mdot$ increases \citep{pejchama}.  

Although a changing bolometric correction is insufficient to explain the accelerating brightening in the $I$ band on its own, it can contribute more significantly to the evolution if $T_{\rm in}$ increases.  However, as a corollary, any implications derived based on the $I$ band flux alone will be highly uncertain, because the bandpass now likely resides on the blue exponential tail of black body and the flux is therefore extremely sensitive to the effective temperature, which we estimate relatively crudely.  For this reason, we show only the bolometric luminosity predicted by our model in Figure~\ref{fig:lc}.

\subsubsection{Response of a star to mass loss}
\label{sec:response}

There are two fairly natural reasons why the value of $T_{\rm in}$ could increase approaching the terminal stages of the merger. First, as the binary loses mass, hotter layers below the surface become exposed.
Using Modules for Experiments in Stellar Astrophysics (MESA) version 8118  \citep{paxton11,paxton13,paxton15}, we perform numerical calculations to explore the appearance of the star subjected to constant hydrodynamic mass loss. The numerical setup mostly follows previous work \citep{passy12}, but we choose the 'Macdonald' equation of state to obtain a smoother evolution.  We evolved a star according to MESA's `1M\_pre\_ms\_to\_wd' test suite and saved a snapshot when the radius reached $10R_{\bigodot}$ after the main sequence. This procedure should roughly recover the structure of the heavier star, which is slightly inconsistent with the assumption of mass-loss coming from the lighter star (Sec.~\ref{sec:ltwo}). However, we do not know from which star is the mass actually coming and the two stars share their surface layer, where the heavier star probably dominates (at least for energy generation).

The saved model was then subject to constant hydrodynamic mass loss using `mass\_change' control. The mass-loss rates we study probe the dynamical response of the star, where the thermal timescale is much longer than the mass-removal timescale. The maximum timestep was set to be smaller than the dynamical timescale of the star to properly resolve the transient pulsations. Although the details of the radius evolution depend sensitively on resolution and other numerical parameters, we find that the evolution of effective temperature is relatively robust. After the removal of mass commences, the effective temperature (and luminosity) of the star rise and  asymptote to nearly-constant value within $\sim 1$ day. These asymptotic values are nearly power-law functions of $\mdot$ with a kink in the dependence at around $10^4$\,K due to hydrogen ionization. Figure~\ref{fig:tsurf} shows the asymptotic effective temperatures as a function of time using the evolution of $\mdot(t)$ that we infer for V1309~Sco. The luminosities follow a similar trend as the temperature, because the fractional change in the radius is relatively small. 

A similar increase in the effective temperatures and luminosities due to dynamical-timescale mass loss was reported previously \citep{passy12,pavlovskii15}. We caution that this behavior differs from the more commonly investigated thermal timescale mass loss, which leads to the dimming of the mass-losing star, because the thermal energy is expended on radial expansion \citep[e.g.][]{gotberg17}. Furthermore, such high mass loss rates push 1D stellar evolution codes such as MESA near or beyond their design limits. Without further verification, our results on effective temperatures and luminosities should be regarded as only qualitative.

Although no color information is available during the last $\sim 200$ days prior to the peak of V1309~Sco, it appears unlikely that the gradual brightening in the first 8 months of 2008 is simply the result of heating of the binary surface without additional reprocessing by an external medium. First, the binary surface is no longer visible, as evidenced by the lack of variability during this time.  In addition, similar gradual brightening phases observed in other stellar merger transients such as M101 OT2015-1 \citep{blagorodnova17} suggest that the effective temperature decreases relative to the progenitor during this phase, implying a much larger photospheric radius.

\subsubsection{Shearing motions within the contact binary}

Alternatively, gas leaving the binary could be pre-heated by the energy dissipated by shearing motions within the binary, due to a moderate loss of co-rotation. Similar possibility together with ejection of hotter layers was also mentioned by \citet{kashi17}. The observed period evolution of V1309~Sco proceeded so rapidly that the cores of the two contact binary components could not maintain synchronicity with the orbit \citep{tylenda11}. This results in a shearing layer between the two stars \citep{tylenda11} with a velocity differential of the order of $\Delta v \sim (1-P/P_0)v_{{\rm orb},0}$ (Fig.~\ref{fig:massloss}), where $P_0$ is the orbital period at the last moment when cores and envelope were corotating (assumed to be the beginning of the OGLE dataset), and $v_{{\rm orb},0} = \sqrt{GM/a_0}$ is the corresponding orbital velocity. Energy can be efficiently transported from the shearing layer to the surface by convection and in steady-state the specific internal energy of the gas will be increased by $\sim \Delta v^2/2$ if adiabatic and radiative losses can be neglected. Our results in Figure~\ref{fig:tsurf} show that this heating can appreciably increase the surface temperature in the last $\sim 100$ days ($\sim 70$ orbits) prior to the dynamical event. This process was likely accompanied by magnetic field generation or amplification, which is of potential interest \citep[e.g.][]{tout08,ohlmann16,schneider16}. In a fraction of similar cases, the magnetic field generated in this way could potentially play a role in launching of jet-like outflows that are observed in some planetary nebulae \citep[e.g.][]{sahai98,bollen17}.

The existence of shear heating within the binary is not necessarily in conflict with the assumption of corotation at \ltwo inherent to our SPH simulations.  If the shearing is restricted to sub-synchronously rotating cores, the small amount of material in the envelope shared by the contact binary (with relatively little inertia) could still be forced to corotate with the orbit.  If matter launched from near \ltwo is not in corotation with the orbit, then tidal torquing might be less efficient and mechanical interaction with the binary could heat the gas and contribute to its outward acceleration.

\section{Discussion and Conclusions}
\label{sec:discussion}

We have argued that the pre-maximum photometric evolution of V1309~Sco observed in detail by OGLE can be explained as a consequence of \ltwo mass loss with progressively increasing mass loss rate. Initially, the resulting stream of gas is transparent and its effects could be seen only in the orbital period evolution, possibly on top of other effects such as the Darwin instability. When the stream becomes semi-transparent, it begins to modify the phased light curve and eventually the double-hump profile arising from mirror symmetry along the binary axis is replaced by single-hump profile (Fig.~\ref{fig:phase}). The mass loss rates estimated from the phased light curves roughly match the simultaneous decay in orbital period (Fig.~\ref{fig:mdot}), although there exist degeneracies between the mass-loss rate, inclination, opacity, and density profile of the outflow. 

As mass-loss rate increases even further, the binary becomes obscured and the observer sees only the spiral collisions in the \ltwo outflow. Extrapolating the runaway trend of $\mdot$ increase, we can reproduce the observed accelerating brightening to the maximum (Fig.~\ref{fig:lc}) provided that the particle injection temperature in the final $\sim 100$\, days increases and raises the radiative efficiency of the outflow. Our results imply that V1309~Sco lost a total of $\approx 0.05\,\msun$ over the last few thousand orbits before the final dynamical mass ejection. The total amount of mass lost is comparable to what hydrodynamical simulations predict for the subsequent dynamical phase \citep{ivanovasci,nandez}. The timescale of \ltwo mass loss is much longer than suggested by contemporary hydrodynamical simulations, for which the initial mass ejection occurs almost instantly and might be followed by a prolonged period of mass loss \citep[e.g.][]{lombardi11,nandez,clayton17}.

What are the prospect of observing the full evolution of a merging binary system similar to V1309~Sco again in the future?  The OGLE data suggests that when \ltwo mass loss sets in, the runaway lasts at most few thousand orbits, corresponding to a few years for short-period binaries. Before this phase, the binary might be relatively inconspicuous in photometric datasets. The frequency of stellar mergers in the Milky Way is roughly once every decade \citep{kochanek14}, thus implying that only a few binaries are undergoing accelerating period change in the entire Milky Way at any given time.  Given this fact, it seems unlikely that a similar event is happening in the {\em Kepler} field, as reported in KIC~9832227 \citep{molnar17}. Furthermore, the reported mass ratio of this system seems too high for the tidal Darwin instability.  Instead, this binary likely falls in the tail of the period change distribution observed in contact binaries, due to additional orbiting bodies in the system, stellar spots, or mass transfer effects \citep{pietrukowicz17}.

For binaries observed from inclinations smaller than $\sim 70^\circ$, the observer's sight line will not intersect the \ltwo mass loss stream. In such cases, the separate contributions to the observed flux from the binary and the shock-heated \ltwo outflow should simply add together.  This leads to the prediction of a gradually rising light curve without a major dips or early onset of asymmetry in the phased light curve.  Unfortunately, a smaller inclination angle also reduces the amplitude of the binary variability, making the discovery of such a system in archival data challenging. For binaries viewed closer to edge-on, the dust formation in the equatorial outflow could potentially hide the central binary and the spiral shocks at very early times. The object would then become optically-visible again only after the release of the more spherical dynamical ejecta. This is similar to the ``Type II ILOT'' of \citet{kashi17}. In systems similar to V1309~Sco, this would happen only for a narrow range of  inclinations $i \gtrsim 85^\circ$, but the actual threshold depends sensitively on the density profile of the \ltwo stream.

The origin of the rising temperature of the gas being injected from $L_{2}$, which from our simulations we find is required to explain the accelerating brightening to the main peak (Fig.~\ref{fig:tsurf}), is unclear.  It might be connected to the mass stripping of the surface layers or additional heating due to internal shears within the binary.  Whatever its nature, our results suggest this heating process operates at least for many tens of orbits prior to the dynamical phase.  We further speculate that the entirety of the runaway process observed in V1309~Sco could be ``self-regulated", in the sense that the evolution $\mdot(t)$ is determined by the expansion of the stellar radius, which is driven by shear heating, which in turn is determined by the integrated history of mass loss and period decay. In this picture, the late-stage self-obscured binary might appear very different than at earlier times (Fig.~\ref{fig:phase}), with strong surface flows/convection, mass ejection and perturbations to the mass loss stream.  This complex behavior is captured only heuristically in our simulations by increasing $T_{\rm in}$, but direct time-dependent multi-dimensional simulations of this phase are unfortunately challenging due to long timescales and complex physical processes (e.g.~convection, radiative transport).  Even the time-steady structure of contact binary envelopes remains an unsolved problem \citep{rucinski15,shu81,kahler04,eggleton06}.

What triggered the ultimate dynamical event in early September 2008?  Our requirements on the mass loss needed to explain the slow brightening phase prior to this point show that the dynamical transition occurs at roughly the point when an order unity fraction of the binary's total original angular momentum has been lost through \ltwo.  We thus hypothesize that reaching this critical point--at which the stellar cores have become sufficiently close to interact directly--was responsible for initiating the final dynamical coalescence and its concomitant mass ejection.

\begin{figure*}
\centering
\includegraphics[width=0.8\textwidth]{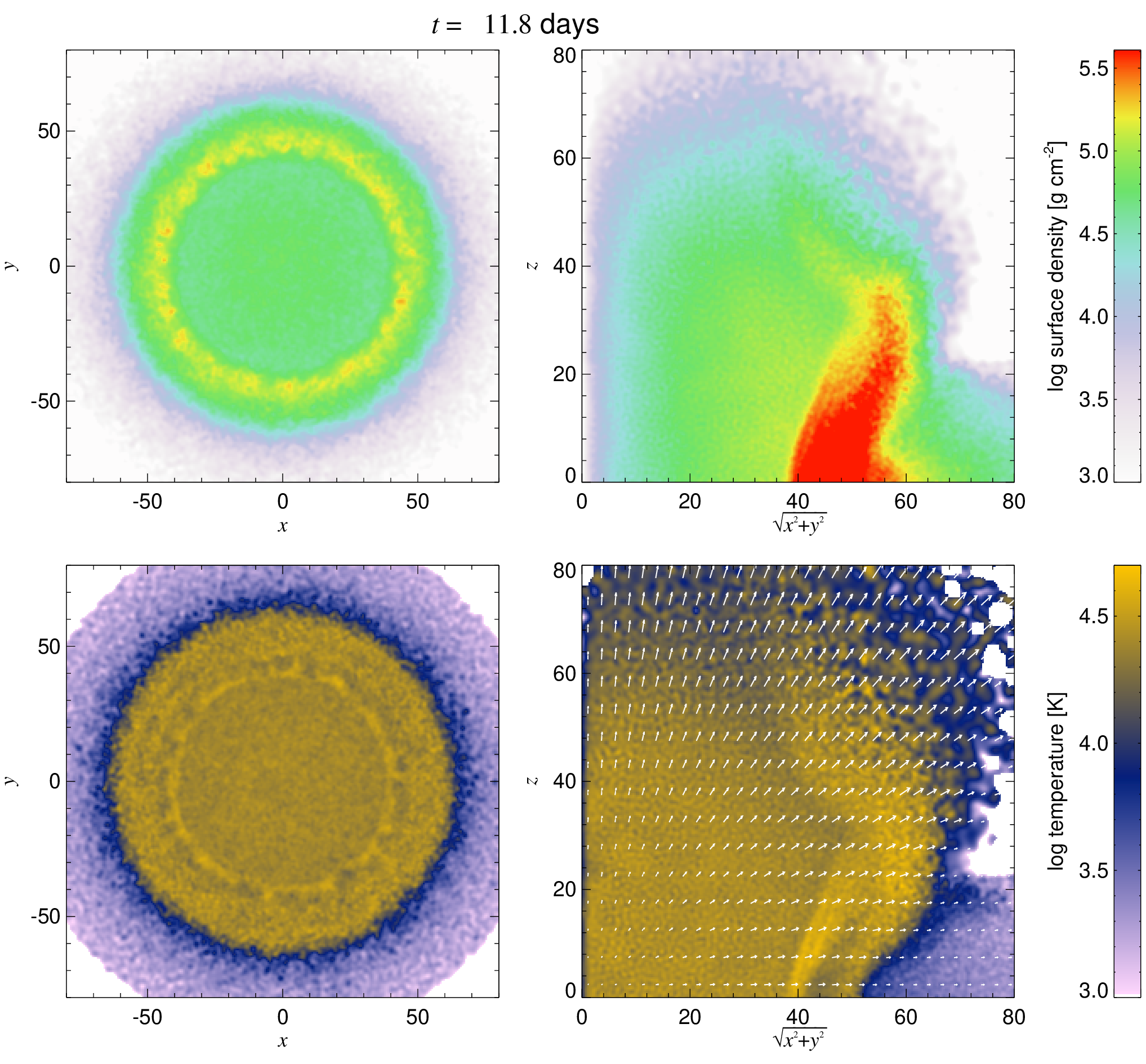}
\caption{Density and temperature structure of the collisions of faster spherical ejecta with pre-existing equatorial \ltwo outflow. The two left panels show density and temperature of particles projected in the orbital plane, while the two right panels show the same quantities in the $\sqrt{x^2+y^2}$--$z$ plane. The initial condition for the equatorial component is the final snapshot from the simulation of gradual brightening of V1309~Sco (inset of Fig.~\ref{fig:lc}). The spherical explosion was initiated by positioning a stationary ball of hot particles in a sphere centered on the binary barycenter similarly to \citet{metzger17}. Gravitational forces and radiative diffusion and cooling were not included in this calculation. The swept-up shell in the equatorial plane fractures into clumps due to hydrodynamical instabilities. Formation of such a clumpy ring can be obtained in different settings, such as a propagation of a jet into a spherical shell \citep{akashi15}. An animated version of this figure is available in the online material.\label{fig:explosion}}
\end{figure*}

Our models show that roughly $0.05\,\msun$ was lost from the binary prior to the dynamical event, an amount of mass comparable to the quantity $0.04$--$0.09\,\msun$ needed to explain the main light curve peak and subsequent plateau \citep{ivanovasci,nandez}.  The mass ejection which occurs promptly during the dynamical phase is expected to be faster and approximately isotropic in geometry \citep{tylenda11}, in which case it will drive a shock through the earlier \ltwo outflow accumulated primarily along the orbital plane.  The shock will soon be engulfed within the more spherical recombining ejecta, with the additional heating from the shocked shell acting to lengthen the duration of the hydrogen recombination plateau \citep{metzger17}, similarly to radioactive nickel decay or magnetar spindown in Type II supernovae \citep{kasen09,sukhbold17}. Conversion of shock kinetic energy to radiation is most efficient when the masses in the two colliding media are similar, as we find appears to be the case for V1309~Sco. We illustrate this scenario qualitatively in Figure~\ref{fig:explosion}, where we initiate a spherical explosion inside the equatorial density structure assembled during the pre-maximum evolution of V1309~Sco.

The collision of spherical fast ejecta with a slower equatorial outflow naturally leads to double-peaked light curves, as observed in many transients in the class similar to V1309~Sco \citep{metzger17} and this suggests that stellar binary mergers commonly lose significant mass well before the dynamical phase. 
If the dynamical event results in a similar total mass ejection as the preceding equatorial \ltwo phase ($\sim 10\%$ of the total binary mass), then the resulting shock interaction can explain the observed range of luminosities and durations of transients from binary interactions.  Depending on the shock energy set by the binary parameters, the collision might not be energetic enough to keep the hydrogen in the fast expanding shell ionized and there would not be a second peak. \citet{metzger17} suggest that this indeed might be the case for V1309~Sco and perhaps also M31 LRN 2015 \citep{kurtenkov15,williams15}. If the shock cannot prevent dust formation, which should occur more readily in long-period binaries with more evolved components, the forming dust will shift the peak of the spectral energy distribution to the infrared and prolong the transient duration. This provides an explanation for some members of the recently identified class of infrared transients known as SPRITEs, which have durations of hundreds of days to years and luminosities in the gap separating classical novae and supernovae \citep{kasliwal17}. It is currently unclear what fraction of the binary mass is ejected in these events, but SPRITEs may represent mergers in which the outcome is a binary surviving on a tighter orbit (the classical definition of a ``common envelope").  Transients identified at optical wavelengths, including V1309~Sco, appear to originate from more compact progenitors in which case most of the envelope is retained, leading to a complete merger.

Modeling transient events in their entirety, including the early slow rise phase studied here, offers a new avenue to explore the progenitor binaries, probe binary stability across a wide range of stellar properties, reduce uncertainties in binary population synthesis models, and provide realistic initial conditions for multi-dimensional numerical simulations of the dynamical phase. 

V1309~Sco remains enshrouded in dust, making direct proof of a single-star remnant challenging to obtain. The bolometric luminosity of the remnant is significantly higher than prior to the outburst \citep{tylenda16}. If the merger remnant is a single star, it is undergoing thermal relaxation.  Its long-term light curve will encode information on the energy and angular momentum injected into the star, as probed by the directly-observed dynamical phase of the merger.  With new data coming from time-domain surveys from ground and space, including Gaia, now is an opportune time to explore the full range of signatures of catastrophic interactions in binary star populations \citep{rucinski01}.

\acknowledgements

O.P. is currently supported by Primus award PRIMUS/SCI/17 from Charles University. In the past, O.P. was supported by program number HST-HF-51327.01-A provided by NASA through a Hubble Fellowship grant from the Space Telescope Science Institute, which is operated by the Association of Universities for Research in Astronomy, Incorporated, under NASA contract NAS5-26555, and in part by the National Science Foundation under Grant No. NSF PHY-1125915 to KITP. B.D.M. acknowledges support from the National Science Foundation (AST-1615084) and NASA Astrophysics Theory Program (NNX16AB30G, NNX17AK43G). We thank Ronald Taam for comments on the early version of the manuscript. We thank the referee for comments that helped to improve the paper.

\begin{deluxetable*}{cp{7cm}p{7cm}}
\tablecaption{Explanation of parameters used in this work\label{tab}}
\tabletypesize{\footnotesize}
\tablehead{
\colhead{Parameter} & \colhead{Explanation} & \colhead{Value or comment}
}
\startdata
$M_1$ &  mass of the lighter star & $0.15\,\msun$ for phased light curve, $\mdot_1 \equiv \mdot$ for gradual rise to maximum\\
$M_2$ &  mass of the heavier star & $1.5\,\msun$\\
$q$   &  mass ratio               & $M_1/M_2$\\
$a$   &  semi-major axis          & $0.03$\,AU for phased light curve, for gradual rise to maximum calculated from Eq.~(\ref{eq:pdot})\\
$P$   &  orbital period           & calculated from Kepler law or Eq.~(\ref{eq:pdot})\\
$v_{\rm orb}$ & orbital velocity & $\sqrt{GM/a}$ \\
$v_{\rm esc}$ & escape velocity & $\sqrt{2GM/a}$ \\
$t$   & time  & \\
$E$   &  orbital energy           & $-GM_1M_2/(2a)$ \\
$\mathcal{E}_i$, $\mathcal{E}_f$  & initial and final energy (sum of potential and kinetic) of a corotating test particle at \ltwo in units of $GM/a$ & Eq.~(\ref{eq:energies})\\
$\mathcal{A}$  & conversion between $\dot{P}/P$ and $\dot{M}/M$, depends only on $q$ \citep{pribulla98}  & typical values $30$--$50$ for $q=0.1$, Eq.~(\ref{eq:pdot})  \\
$f$ &  additional angular momentum extracted from the binary orbit by a particle released from \ltwo and moving to infinity  & assumed mostly $1.2$,  Eq.~(\ref{eq:pdot})  \\
$\varepsilon$ & initial width of \ltwo stream & $0.01$ in SPH calculations, roughly proportional to $c_S/v_{\rm orb}$ \citep{shu79}\\
$\rho_{\rm center}$ & density at the center of \ltwo stream & Eq.~(\ref{eq:rho_center})\\
$\psi$  & perpendicular distance from the \ltwo stream center  & in units of $a$, Eq.~(\ref{eq:rho_psi}) \\
$T_{\rm binary}$  & photospheric temperature of the binary & $4500$\,K, Eq.~(\ref{eq:t_binary}) \\
$I$  & intensity along a ray& Eq.~(\ref{eq:ray})\\
$s$  & distance along a ray& Eq.~(\ref{eq:ray})\\
$\kappa$ & opacity & Eq.~(\ref{eq:ray}) \\
$i$ & inclination & preferred value is $84^\circ$ given the assumed density profile of the \ltwo stream \\
$\varphi$ & orbital phase & obtained by integrating $1/P$\\
$t_0$  & time of singularity in $\mdot$ & corresponds roughly to the main peak, Eq.~(\ref{eq:mdot})\\
$\delta$ & exponent of runaway of $\mdot$ &  ``best-fit'' value $3$, Eq.~(\ref{eq:mdot})\\
$T_{\rm in}$  &  injection temperature of particles at \ltwo &  Eq.~(\ref{eq:tin})   \\
$T_{{\rm in},0}$  & initial  injection temperature of particles at \ltwo &  equal to $T_{\rm binary}$, Eq.~(\ref{eq:tin})   \\
$\mdot_T$   & critical $\mdot$ above which $T_{\rm in}$ begins to increase &  ``best-fit'' value of $10^{-2}\,\myr$, Eq.~(\ref{eq:tin})\\
$\gamma$ & exponent of runaway of $T_{\rm in}$  &  ``best-fit'' value $0.5$, Eq.~(\ref{eq:tin})\\
$b_j$, $c_j$ & coefficients of Fourier expansion of phased light curve & Sec.~\ref{sec:phased_results}\\
\enddata
\end{deluxetable*}

\end{document}